# Principal Component Analysis for Experiments

Tomokazu Konishi[1,*]

[1]Fuculty of Bioresource Sciences, Akita Prefectural University, Shimoshinjyo Nakano Akita, Japan.

**ABSTRACT**

**Motivation:** Although principal component analysis is frequently applied to reduce the dimensionality of matrix data, the method is sensitive to noise and bias and has difficulty with comparability and interpretation. These issues are addressed by improving the fidelity to the study design. Principal axes and the components for variables are found through the arrangement of the training data set, and the centers of data are found according to the design. By using both the axes and the center, components for an observation that belong to various studies can be separately estimated. Both of the components for variables and observations are scaled to a unit length, which enables relationships to be seen between them.

**Results:** Analyses in transcriptome studies showed an improvement in the separation of experimental groups and in robustness to bias and noise. Unknown samples were appropriately classified on pre-determined axes. These axes well reflected the study design, and this facilitated the interpretation. Together, the introduced concepts resulted in improved generality and objectivity in the analytical results, with the ability to locate hidden structures in the data.

## 1 INTRODUCTION

The application of sensitive detection systems to separation methods has led to recent technologies that can identify and measure various substances for quantitative omics studies. Examples include chromatograph systems coupled with mass spectrography, used in metabolomics, and arrayed nucleotides with fluorescent probes, used for transcriptome. In addition to determining the levels of many substances, these comprehensive approaches enable the understanding of complex biological phenomena as a physical expression of genomic information. Hence, omics studies are advantageous for obtaining the whole picture of the phenomena, and they have rapidly become popular. Not only can these technologies be used to observe an event in specially designed experiments, they can be used for examinations or surveys in medical research.

The analysis of the resulting data and the integration of the obtained knowledge from omics studies have become important issues. The data are provided as a matrix with observations by variables, and these data have two characteristics. First is the number of variables corresponding to the measured items; this number is quite large. Second, the samples to be observed have intricate interrelationships that originate from the study design. The huge size and complexity make the data difficult to understand using direct methods.

To help facilitate the understanding of the matrix data, researchers have had success in applying hierarchical clustering (Eisen et al., 1998). In this method, the distance is defined between a pair of objects, and the distances are summed for groups of objects. Then, the method finds the closest pair of objects or groups, records the distance, and unites the pair to form a new group. This process is repeated until a group is found that includes all of the objects. A record of the distances is presented in a dendrogram, which shows the cluster structure of closely-related objects or groups. However, hierarchical clustering has a limited capability for comparing and integrating dendrograms obtained from different studies. Also, it is limited in its objectivity, since the results strongly depend on an arbitrary choice for the definition of distance, chosen from some tens of possible combinations. As the observations and variables are independently analyzed, their relationships are difficult to find. Obviously, these present critical limitations to a systematic understanding of the data.

Principal component analysis (PCA) (Jolliffe, 2002; Jackson, 2005) may provide possible alternative approaches to understanding quantitative omics data. Indeed, PCA and its variations have been applied to transcriptomic microarray data (for reviews, see Alter, 2007; Ringner, 2008). PCA approximates many interrelated objects by extracting their data differences into several independent principal components (PCs). This reduces the dimensionality of the matrix data, and simplifies subsequent interpretations. Fortunately, many of the variables in omics data are actually correlated, since they reflect items that are controlled by a few common cellular factors (Konishi, 2005). Hence, PCA is a promising approach for obtaining a proper solution for the data.

However, there are some disadvantages to the PCA method as applied to quantitative omics data. First, since the method does not consider weights and relationships among the observations, it has only a limited capability for responding to a study design which has been planned to manage noise and bias. Hence, this rigidity reduces not only the fidelity to the data structure but also decreases robustness to noise and bias. Interestingly, both noise and bias have been addressed through two distinct approaches: sophisticated calculation methods and improved fidelity to the data structure (described below).

Noise is mainly derived from the individual differences in the samples as well as from measurement error, and it reduces both sensitivity and accuracy in the location of the axes. To reduce the effect of outliers, a robust calculation method has been applied to microarray data (Liu et al., 2003). Alternatively, Jeffers (1962) applied PCA to the means of repeated observations from each group, and concentrated them to intra-group differences to reflect

[*]To whom correspondence should be addressed.

*T. Konishi*the experimental design. Likewise, the removal of items that show large inter-group variances (de Haan, 2007) could reduce the effect of noise.

Bias is derived from various sources, including from experimental errors or by chance in sampling, and it results in incorrect directions of the PCs; indeed, Skinner et al. (1986) identified the instability of PCA in the context of sampling design. LASSO and related methodologies (Jolliffe et al., 2003; Zou et al., 2006) reduce the effects of bias by ignoring smaller differences between samples. However, the data metrics are lost by these methodologies, creating additional problems with the fidelity and objectivity of the analytical results. Konishi and Rao (1992) introduced a methodology for data that contains families with differing numbers of siblings. This approach could further improve the fidelity to the data structure compared to the use of the group means. However, bias could also derive from the experimental design itself. Particularly, the number of families may be biased; for example, toxicology experiments may be composed of observations that are biased toward particular families of chemicals. This resulting bias would affect both the directions and the magnitude of the PCs, as will be observed later in the RESULTS (Section 3).

Another critical limitation of PCA is the generality of the resulting components. The PCs have limited comparability among different studies (Jolliffe, 2002; Jackson, 2005). Rather, the method has been used to estimate possible interpretations that are valid only within a study (Shaw, 2003). This is a fundamental limitation that is based on the definition of PCs, as follows. Let $X$ be an observation-by-variable matrix, containing variables with zero means. By singular value decomposition, the matrix is factorized into one diagonal and two unitary matrices, $X = UDV$; the unitary matrices $U$ and $V$ define the directions of the principal axes, while the diagonal matrix $D$ records the magnitude of the data differences as the ordered singular values. The PCs for observations are defined as $Y = XV$ (Jolliffe, 2002; Jackson, 2005); obviously, this equation presents a rotation of $X$ according to $V$, so that variations in $X$ are aligned to the directions of the orthogonal axes. For example, the first principal axis indicates the direction representing the largest variation among the observations. Accordingly, the magnitude of $Y$ depends on the number of variables; the magnitude increases with more variables. Also, since $V$ is found by the decomposition of $X$, different combinations of observations that form $X$ lead to different directions of the rotation. Since they use different scales and directions, the PCs found in different studies cannot be compared. Indeed, comparisons or integrations of different observations from multiple studies have only been attempted when likely directions could be expected (Alter et al., 2003). Although the integration of PCs by multi-block PCA methodologies (Westerhuis, 1998) have been applied to omics data (Spicker, 2008), these methodologies were intended to integrate different measured items within the same study and not observations taken from different studies.

Needless to say, natural science is based on the organization and integration of knowledge obtained from many studies. In particular, omics studies aim to understand biological phenomena in a systematic way, and they rely heavily on the quantitative measurements of substances. To accomplish the purposes of these studies, it is best to maintain the metrics of the data, as they are compatible or closely related to the physicochemical framework. Flexible frameworks, such as those applied using a LASSO-related approach to transcriptome analysis (Van Deun et al., 2011), introduce conditions that are specific to the data being analyzed; hence, comparisons of the results to other studies are difficult.

In this article, new concepts are introduced to solve the problems of noise and bias, poor fidelity to the data structure, and low generality. First, processes to find and apply $V$ are intentionally disconnected; the unitary matrix is found in the arranged training data $T$ and is applied to each observation. This enables the principal axes to be shared between studies. This ability to share is useful as it allows comparisons of biological effects as well as the categorization of unknown samples into known groups (see RESULTS, Section 3). Additionally, in the arrangement of $T$, noise can be reduced using repeated measurements as well as by removing unstable items. Still, we can observe the level of noise through fluctuations of the calculated PCs for observations. Also, biases in families could be adjusted by selecting a proper set of observation groups. Second, the center of the rotation is determined in accordance with the experimental design. The de facto standard of the center is item-wise means; however, these means are not always the best choice since they are sensitive to biases in the groups. Instead, many experiments have a control group that is suitable as the center of the rotation. Such a control group, measured in each experiment, is used to provide a standard that cancels trivial differences among experiments. Third, scaling of PCs is introduced to estimate the mean contribution of variables. This enables a comparison of the magnitude of PCs that have been found on different numbers of stable items. Additionally, instead of using the loadings, PCs for variables will be estimated and then scaled. This reflects the singular values to the loadings, and facilitates the comparison of PCs from observations and variables. Through such comparisons, it becomes clear how particular items contribute to the characteristics of the observations.

These concepts were tested in two transcriptome studies. The effects of data noise and sample bias were observed in a separation of the sample groups. Also, the appropriateness of the scaling was checked based on the constancy of the PCs' ranges between different numbers of items. The possibility of sharing axes, which enables the use of a common framework among experiments, will also be discussed.

## 2 METHODS

**2.1 Preparation and pretreatment of the data matrix**

Two transcriptome studies that investigated several experimental groups using repeated measurements were found in the Gene Expression Omnibus (Barrett et al., 2009). These studies were a time course in mammary gland development (GSE8191) that was conducted by Anderson et al. (2007), and toxicology data obtained for some carcinogens (GSE5509) that was conducted by Spicker et al. (2008). Perfect match sample data were parametrically normalized to find the z-scores, and the gene expression levels were estimated as the trimmed means of the z-scores (Konishi et al., 2008) using the SuperNORM data service (Skylight Biotech Inc., Akita). The normalized perfect match data and summarized gene data are available in the Gene Expression Omnibus (GSE31375). The significance in the expressional changes was tested using a two-way analysis of variance (ANOVA) on the normalized perfect match data, assuming a linear relationship, in which (data difference) = (probe sensitivity) + (group effect). A threshold of 0.005 was used to find the positive genes (Konishi, 2011).

**2.2 Determine the reference of experiment as the center of data**

The matrix of the expression levels is subtracted using the reference data,

$$X = \begin{pmatrix} s_{11} & \cdots & s_{1m} \\ \vdots & \ddots & \vdots \\ s_{n1} & \cdots & s_{nm} \end{pmatrix} - \begin{pmatrix} r_1 & \cdots & r_m \\ \vdots & \ddots & \vdots \\ r_1 & \cdots & r_m \end{pmatrix},$$

where $s$ is the normalized sample data, $r$ is the reference found for each variable, and n and m are the numbers of observations and variables, respectively. The reference, which determines the origin of the PCs, were found as the item-wise means of all the samples for the time-course data, or of the control group for the toxicology data. Missing or removed objects in $X$ were replaced with zero, which means there was no difference from the reference.

### 2.3 Hierarchical clustering of data

Hierarchical clustering was performed on X using the *dist* and *hclust* functions in R (R Development Core Team, 2012). The Euclidian distance was found between pair of objects and summarized using Ward's distances. Only genes that were positive on the ANOVA test were used for the analysis.

### 2.4 Finding the orthogonal axes of PCA in training data

The principal axes were identified in a training data matrix, $T$, which was intentionally prepared. Unless otherwise stated, $T$ is the set of the means of observations in each group, for example, each point in the time-course data or each chemical in the toxicology data, with the reference value subtracted. Next, the training data was decomposed as $T = U_T D_T V_T'$. The PCs for the observations were defined as $Y = XV_T$, and likewise, the PCs for the variables were defined as $Y_v = T'U_T$. Since these are rotations of $T$ and $T'$, the original metrics of the data are maintained. Additionally, for $T' = V_T D_T U_T'$, $Y_v = V_T D_T U_T' U_T = V_T D_T$; therefore, the derivation of $Y_v$ could also be considered as a reflection of singular values to the loadings of $Y$, i.e., $V_T$.

In the examples presented in this article, the PCs were estimated under various conditions. As the directions of the axes were randomly determined, they were unified among the conditions by simultaneously reversing the positive and negative signs in the corresponding columns of $Y$ and $Y_v$.

### 2.5 Scaling of the principal components

In $X$, the number of functional items, $m_f$, would differ between the observations due to missing data and/or the removal of insignificant variables, as follows. Suppose $m_f$ variables in $X$ are randomly selected to obtain $X_f$, and the variables have zero means. When $X_f$ is decomposed to $X_f = UDV$, any differences in the selection will not have much affect on the ratios among the $i$th diagonal elements of $D$, $d_i$ (this will become obvious in Figures 1C and 1D). Since $D$ records the scales of $X_f$, the sum square of $d_i$ is equal to that of the elements of $X_f$, hence $\sum_i d_i^2 = \sum^{m_f} \sum^n x^2$. Therefore, $\sum_i d_i^2 \propto m_f$. As the ratios among the $d_i$ can be supposed to be constant, $d_i \propto m_f^{1/2}$. Since $Y = UD$, the PCs for the observations are proportional to the square root of $m_f$.

To improve the comparability of the PCs, their unit length should not be affected by $m_f$. Indeed, differences in $m_f$ could become larger in observations of multiple studies, since the measured items may have some differences. Hence, to maintain the magnitude of the PCs regardless of the size of $m_f$, they are scaled as $Z = m_f^{-1/2} Y$. Also, the PCs for variables are scaled as $Z_v = n_T^{-1/2} Y_v$, where $n_T$ is the number of observations in the matrix for which the principal axes were found; this is identical to the "block scaling" of PCs used by Westerhuis (1998). $Z$ and $Z_v$ indicate the average contribution of each variable or observation, respectively. As the scaling does not cancel $D$, the ratios between the axes will be retained.

In Figures 1B-D, fluctuations of observations were found as the average of the standard deviations (SD); in each group, the SD were estimated in the Euclidean distances of sPC1 and sPC2, and the average was estimated by the root mean square of the SD.

### 2.6 Overlying presentation for observations and variables

As is obvious from their definitions, $Y$ and $Y_v$ are closely related and share some characteristics; indeed, in the corresponding PCs that were estimated directly from $X$, the sum squares will be identical, as follows. Since $Y = XV = UDV'V = UD$, the sum square of the $i$th PC is $\Sigma(d_i u_i)^2 = d_i^2 \Sigma(u_i)^2 = d_i^2$, where $u_i$ is the $i$th row of $U$. Likewise, the sum square of the $i$th PC for variables, $Y_v = V_T D_T$, is also $d_i^2$.

Similarly, the root mean squares of the corresponding scaled PCs (sPCs) are identical; since $Z = UD\ n^{-1/2}$, the root mean square of the $i$th sPCs is $(\Sigma(d_i u_i n^{-1/2})^2 /m)^{1/2} = d_i\ n^{-1/2} m^{-1/2}$, which is also found to be that of $Z_v$. This is the mean Euclidean distance of the $i$th sPCs from the origin. Based on this characteristic, a biplot-like presentation was made by overlaying the sPCs for the observations on those for the variables, using identical scales for the axes.

### 2.7 Robust calculations

The effects of using a robust alternative to singular value decomposition algorithms were observed using two different functions of R, the *robustSvd* function (Liu et al., 2003) in the *pcaMethods* library and the *PcaHubert* function (Hubert et al., 2005) in the *rrcov* library. The sPCs were also estimated using the resulting matrices.

## 3 RESULTS

### 3.1 Improvement in separation of observation groups

Since fluctuations due to noise can mask the relationship between groups, a reduction in noise was attempted by using a training data set or by focusing on ANOVA-positive genes; these were investigated in regard to the separation of groups in mammary gland development data (Anderson et al., 2007). In this time-course experiment, ten groups of time points were measured: pregnancy (groups 0 to 5), lactation (groups 6 to 8), and involution (group 9). The sPCs for the observations of the first and second axes are presented in Figures 1A-1C. As the scores were calculated for each observation, each sample's individual differences appeared as fluctuations in the sPCs. Compared with the original method that determined the axes for the full data matrix $X$ (Figure 1A), the separation of groups was improved when the axes were found for ANOVA-positive genes (Figure 1B) or for the training data (Figure 1C); the observations for each group were located closer to each other, creating better separation between the groups. The smaller fluctuations enabled a clearer view of the data structure in Figures 1B and 1C; the developmental stages of the mammary gland appear in the figures as a straight line horizontally along the first axis, and the involution appears vertically along the second axis. Additionally, while the axes in Figure 1B were calculated with only half of the number of genes as in Figure 1A or 1C, the range of the components remained the same, demonstrating that the scaling of the PCs worked well. On the other hand, robust versions of the calculations did not improve the group separations (additional Figure 1); these separations were degraded whether all the genes were used or only the ANOVA-positive genes. A cause of this inefficiency in the robust algorithms could be that individual differences, which may not appear as outliers in particular genes but instead as smaller differences in many genes, were a primary contributor to data noise.

To observe the characteristics of the methodologies, the relationships between observations were analyzed using hierarchical clus-tering (Figure 1D). Although many of the observations on the same time points were allocated to be adjoining in order to form some clusters, the relationships between the clusters did not appear in the phylogenic tree; for example, it was difficult to estimate the relationships between groups 3 and 7 or 9. This was because the method presents distances only for the closest pairs, and distances for other combinations cannot be estimated from the phylogenic tree.

### 3.2 Robustness to individual differences

The effects of using training data were further investigated on the sPCs for the variables. Figures 2A and 2B show the first and second sPCs determined by the original method and by using the training data, respectively. In Figure 2A, there are two distinctive clusters that are separate from the large aggregation at the center of the graph. These clusters have high/low sPC2 scores, and remained closely connected to the corresponding clusters in Figure 2B (CL1 and CL2, shown in green). However, in Figure 2B, cluster CL2 was located inside of the center aggregation.

Expressional levels were observed for genes in both clusters. In cluster CL1, they were changed rather uniformly (Figure 2C), as consistent changes at the involution stage may have contributed the sPC2 of CL1 genes. In contrast, the genes in cluster CL2 showed larger within-group variations (Figure 2D). In the original method, the outliers may have unintentionally provided higher sPC2 to the CL2 genes (Figure 2A). However, the effect of the outlying variables may have been reduced by taking sample means in the training data, thereby locating the cluster closer to the center in Figure 2B.

### 3.3 Robustness to bias toward a group

The effects of training data were further studied using the microarray data of Spicker et al. (2008), which measured three carcinogens (groups 1, 3, and 5), three nontoxic chemicals (groups 2, 4, 6), and mock control using five biological replicates. The training data was set using mean data for all the groups, and only ANOVA-positive genes were used for PCA; the first axis separated the toxic chemicals from the nontoxic ones, and the second axis separated the toxic groups (Figure 3A).

In addition to this training data, two alternative sets were prepared. One alternative training data set was used to simulate bias in families of groups; for one of the carcinogens (group 3), all of the observations were replaced to the sample mean. Hence, the training data contained six group means and five observations for group 3 samples. In the artificially-biased training data, the first axis gave the group 3 samples higher scores; additionally, the second axis separated the group 3 samples (Figure 3B). It is obvious that the bias had unnecessarily increased the importance of group 3, by emphasizing rather trivial differences between the samples.

### 3.4 Classifying unknown samples by applying previously determined $V_T$

To investigate the application of the previously determined or even shared $V_T$, another alternative training data was used to estimate the PCs of unknown samples. The training data was prepared using all sample means except those from group 3. The obtained first axis successfully separated the group 3 samples from the nontoxic chemicals, and the second axis separated the toxic chemicals (Figure 3C).

### 3.5 Relationships between observations and variables

The sPCs for the variables found for $T$ were overlaid on those for the observations of the toxicology data. They could be presented on the same axes, although the sPCs for the observations, which were expected to show characteristics of the axes, appeared near the center (Figure 3D). This convergence will also occur when overlaying the sPCs for the mammary gland development study data (compare the ranges of Figures 1D and 2B). This is reasonable since the number of variables is much larger than that of the observations, while the average distances from the origin are identical (METHODS, Section 2.6).

The relationships of the sPCs for variables and observations were further observed through the functions of the genes, which showed the highest scores at sPC1 for the variables. The annotation key words given to the genes were collected, and the frequencies were counted (Table 1). In the mammary gland development study, sPC1 for the observations showed the direction of the time course (Figure 1C), and genes with a high positive sPC1 (Figure 2B) increased their expression levels along the time course. A frequency analysis for the annotation keywords for the genes showed that "cholesterol biosynthesis" and other related words were the most significant (Table 1). Also in the toxicology study, sPC1 distinguished carcinogen groups by giving them negative scores (Figure 3D). Among the genes with negative PC1, which increased their levels by carcinogen treatment, keywords such as "DNA damage" and "apoptosis" appeared as the most significant (Table 1).

## 4 DISCUSSION

Enhancements to principal component analysis were introduced in this paper. These enhancements include separating the identification and application of the unitary matrix $V_T$, developing options in the training data and the reference, and scaling the components (a typical workflow is presented in Additional Figure 4). The use of sample means for the training data improved the resolution between groups (Figure 1); the resulting axes of PCs became even self-explanatory (compare Figures 1A to 1B and 1C), drastically increasing the objectivity of the interpretation. Also, robustness was achieved in the determination of groups of items characteristic to an axis (Figure 2). This improvement may have been derived by reducing noise effects; indeed, a similar improvement was observed by focusing on ANOVA-positive genes (Haan et al., 2007 and Figure 1B). In a practical sense, the use of sample means and the selection of items could be applied simultaneously. However, it should be noted that biases such as those artificially simulated in Figure 3B can only be solved by using training data and by selecting appropriate groups of observations (Figures 3A and 3C).

The introduced concepts also improved the generality of the PCs. In the original method, both the scale and direction of a PC are valid only within the study, forming a closed intelligent framework; therefore, PCs as well as interpretations for the PCs are difficult to integrate between multiple studies. However, by applying $V_T$ to various experiments, observations obtained from different experiments can be presented in the same axes as for the training data, if many of the items are commonly measured. The effects of trivial conditional differences between studies, such as caused by laboratory temperatures or batches of cell cultures, could be canceled using appropriate references found in each experiment. In addition, the scaling of the components could unify the unit of distance. By sharing the axes and the unit distance, studies can share a common framework. Such inter-study comparisons are especially beneficial in diagnosis and toxicology applications, enabling classification of subjected samples having unknown properties (Figure 3C). Also, we can compare responses found in particular conditions by swapping the axes between studies; this compatibility is advantageous for achieving a deeper understanding of experiments. The allocation of an unknown sample or swapping the observational perspective cannot be expected in hierarchical clustering, which requires recalculation of whole objects without a guarantee that the relationships observed among the known samples will reappear.

To observe the relationships between the items and the samples, overlaying of the sPCs for the observations and variables was introduced as a scaled version of a biplot. This improved the biplot's objectivity. According to its definition (Jolliffe, 2002; Jackson, 2005), the original biplots present *cUD* and (1-*c*)*DV'*, where $0<c<1$; hence, the two terms represent the singular values with an arbitrary proportion. By presenting the sPCs for the observations and variables together, the information for the scale fully appears in both terms; thus, differences in the magnitudes of the axes will become quantitatively obvious. Also, although the ranges of $Y$ and $Y_v$ could differ substantially, $Z$ and $Z_v$ were presented in identical axes (Figure 3D). In practice, however, the presentation for the observations could be adjusted by multiplying $Z$ to some extent, leading to improved legibility of the plot.

On the scaled biplot, items that are typical for certain observation groups will become obvious; such items can be used as indicators for the group. As omics technologies measure subjects comprehensively, they could be rather costly. Measuring fewer items using a simplified methodology, such as enzyme-linked immunosorbent assays for metabolome, could provide a reasonable solution. The capability for such alternatives can be checked by comparing their sPC1 for observations with the original omics data. The sPC1 could be used as the indicator that assembles the selected items. By measuring plural items, the sPC1 will have better sensitivity and robustness than any single item.

The relationships between the items and observation groups are quite important for improved understanding of the data. These relationships became obvious in the enhanced version of PCA; for example, genes that contributed to the separation of group 9 in the sPCs for observations (Figure 1C) could be found in the sPCs for variables (Figure 2B). Indeed, the outlying genes in the second axis showed a common expression pattern (Figure 2C). Also, the sPC1 for observations showed the development of the mammary gland (Figures 1B and 1C) or the effect of carcinogens (Figure 3A), and the genes that showed distinguishing scores on the axes possessed annotation keywords that were reasonably expected for the experiments (Table 1). As discussed, this evaluation of $Z_v$ according to $Z$ actually helped distinguish the characteristic items for particular conditions. On the other hand, the characteristics of observation groups will appear in such correlated items, and determining these items will facilitate a deeper understanding of the biological conditions studied in the experiment. As seen above, the sPCs for variables and observations have complementary information. Such interrelationships may be based on the mathematical derivation of PCs; hence, they are difficult to determine in hierarchical clustering which analyzes observations (Figure 1D) and variables (Additional Figure 1) separately.

Although improved fidelity to experimental designs is achieved through the selection of samples and items for the training data, arbitrary selections should be avoided. For example, repeated random trials on the selections or the bootstrapping of items and observations would result in a better fit to some conclusions, but such applications would spoil the objectivity of the analysis. In practice, the selections of samples would be self-apparent from the study design. Of course, the design and the obtained data structure will differ to some extent, and sometimes errors or undesired biases can cause serious departures. To respond to such differences and to maintain fidelity to reality, it would be beneficial to consider restructuring the groups as well as ignoring some items. This would lead to a better interpretation of the data. Alteration of the selections should be limited for cases that have clear and accountable reasons in view of fidelity to the data structure.

In this article, the weights of the items were not adjusted. Scaling the variables of $X$ is frequently performed in PCA, and this can unify the weights of the items (Jolliffe, 2002). Here it was not performed, although preparing the training data can be regarded as a unification or adjustment of the weights of observations. This is partly because such alteration will change the metrics of the original data from distance to correlation. Since correlation lacks linearity, it is difficult to compare different experiments. Additionally, as measurements in omics studies are often performed on a single measurement system, items may share a certain level of technical noise. Hence, the scaling of items may enlarge the noise level of relatively stable items. Actually, the separations found in the mammary gland data became worse by pre-scaling the variables (additional Figure 2A and 2B; these correspond to Figures 1A and 1B, respectively).

In this article, multi-block PCA (Westerhuis, 1998) was not considered for the following reasons. If the data are obtained by several different techniques, they could be combined to form a single matrix prior to the calculations. In this combination, the weight of each item could be adjusted by multiplying the item with a certain compensation constant. In contrast, multi-block models assume an equal weight for the blocks, even if one block is full transcriptome data and the other is a few biochemical measurements. Also, since the multi-block models repeatedly scale the components according to several perspectives, the final metrics are rather complex and thus are difficult to verify or interpret. Furthermore, loadings are found in each of the blocks and cannot be combined; hence, sharing the axes between studies or comparing the variables and observations are precluded.

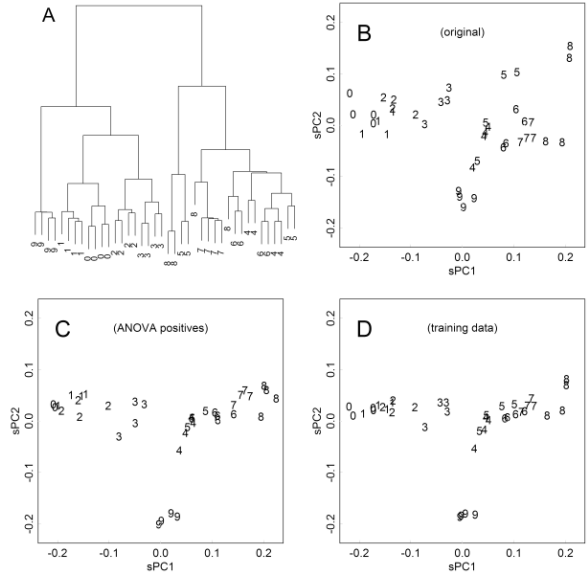

**Fig. 1.** Relationships between the observations in the time-course experiment. Data were obtained for mammary gland development (Anderson et al., 2007). Groups 0 to 5: days 1 to 19 of pregnancy; groups 6 to 8: days 1 to 9 of lactation, group 9: day 2 of involution. **A.** Results of the original PCA. The axes were found with all samples of 12,487 genes. The scaled principal components (sPCs) of the first and second axes are shown. **B.** Axes were found in 5,892 ANOVA-positive genes. **C.** Axes were found in the training data, prepared using each group's sample means of the 12,487 genes. For Figures 1A, 1B, and 1C, the average standard deviation, which would present degrees of fluctuation, were 0.052, 0.029, and 0.028, respectively. **D.** Results of hierarchical clustering.

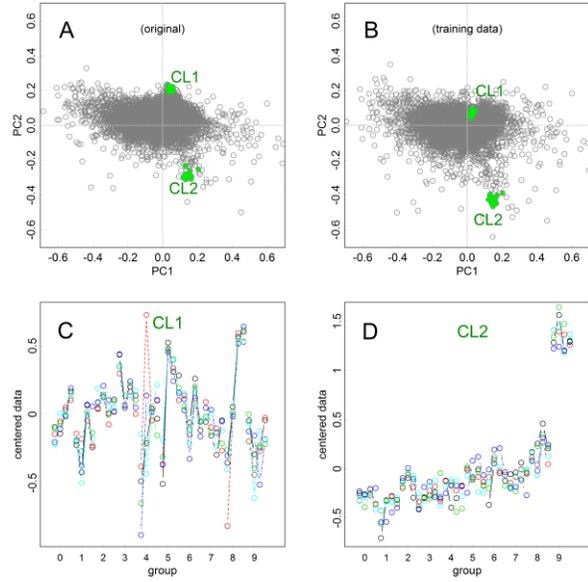

**Fig. 2.** Effect of the training data on the scaled principal components for variables. **A.** Results of the original method. The two clusters, CL1 and CL2, are shown in green. This corresponds to Figure 1A. **B.** Axes were found in the training data, corresponding to Figure 1C. **C.** Expression levels of five genes randomly selected from CL1. **D.** Expression levels selected from CL2, showing higher within-group variations.

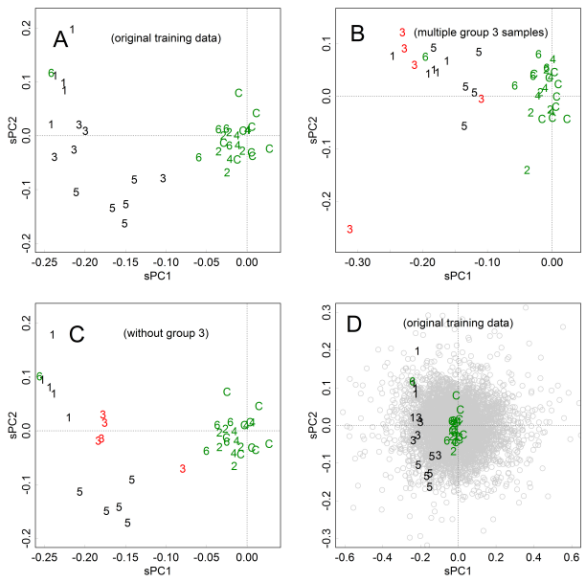

**Fig. 3.** Scaled principal components for samples in toxicology data. Data was measured by Spicker et al. (2008) and found in various training data sets. Groups 1, 3, and 5 were those for carcinogens; groups 2, 4, 6 were those for nontoxic compounds; and C was the mock control. **A.** Training data was prepared by using all of the groups' sample means. **B.** To mimic bias, the sample means of group 3 were replaced with all of the samples to prepare the training data. **C.** To rehearse classifying unknown samples, the training data were prepared without group 3, and the data of group 3 were subjected to PCA. **D.** An example of a biplot-like presentation. The scaled principal components of the training data for variables and observations are presented using identical axes.

**Table 1.** Key words found with the smallest p-values in PC1 of the studies

| key words | chip | selected | p-value |
|---|---|---|---|
| **Mammary gland development** | | | |
| cholesterol biosynthetic process | 33 | 10 | 1.3.E-06 |
| sterol biosynthetic process | 42 | 10 | 1.1.E-05 |
| steroid biosynthetic process | 62 | 12 | 1.1.E-05 |
| isoprenoid biosynthetic process | 16 | 5 | 5.3.E-04 |
| G1/S transition of mitotic cell cycle | 35 | 7 | 6.3.E-04 |
| **Toxicology** | | | |
| regulation of DNA damage response, signal transduction by p53 class mediator | 12 | 5 | 6.3.E-10 |
| regulation of anti-apoptosis | 64 | 7 | 2.1.E-09 |
| regulation of cell cycle | 157 | 9 | 2.7.E-09 |
| apoptosis | 1313 | 19 | 3.9.E-08 |
| response to organic cyclic substance | 321 | 10 | 1.0.E-07 |

In the mammary gland development study, 500 genes that showed the largest PC1 were selected, and the keywords for the Gene Ontology Biological Process were found. The number of words among the whole "chip" contents and the "selected" genes are shown. P-values were calculated according to the binominal model (Konishi et al., 2009). In the toxicology data, 100 genes that showed the smallest PC1 were selected.

## ACKNOWLEDGEMENTS

The author wishes to thank Dr. Shohab Youssefian for encouraging the work.